# Fingering patterns in hierarchical porous media

Si Suo[1], Mingchao Liu[2, 3], and Yixiang Gan[1, *]

[1] School of Civil Engineering, The University of Sydney, NSW 2006, Australia

[2] Mathematical Institute, University of Oxford, Oxford OX2 6GG, United Kingdom

[3] Department of Engineering Mechanics, CNMM & AML, Tsinghua University, Beijing 100084, China

[*] Corresponding author: yixiang.gan@sydney.edu.au (Y. Gan)

**Abstract:** Porous media with hierarchical structures are commonly encountered in both natural and synthetic materials, e.g., fractured rock formations, porous electrodes and fibrous materials, which generally consist of two or more distinguishable levels of pore structure with different characteristic lengths. The multiphase flow behaviours in hierarchical porous media have remained elusive. In this study, we investigate the influences of hierarchical structures in porous media on the dynamics of immiscible fingering during fluid-fluid displacement. Divided by the breakthrough, such displacement process includes pre- and post-breakthrough stages during which the fingering evolution is dominated by viscous and capillary effects, respectively. Through conducting a series of numerical simulations, we found that the immiscible fingering can be suppressed due to the existence of secondary porous structures. To characterise the fingering dynamics in hierarchical porous media, a *phase diagram*, which describes the switch among the three fingering modes (the suppressing, crossover and dendrite mode), is constructed by introducing a scaling parameter, i.e., the ratio of time scales considering the combined effect of characteristic pore sizes and wettability. The findings presented in this work provide a basis for further research on the application of hierarchical porous media for controlling immiscible fingerings.

# 1. Introduction

Fingering phenomena are commonly observed during fluid-fluid displacement processes in porous media. Understanding the fingering dynamics is beneficial to control the fingering patterns, and is further of relevance to many engineering fields, such as $CO_2$ sequestration and storage [1-3], enhanced oil recovery [4], agricultural irrigation [5, 6], and liquid-gas exchange in polymer-electrolyte fuel cells [7, 8]. Porous media encountered in natural and synthetic materials usually contains heterogeneities of properties (e.g., wettability and surface features) and topological structures (e.g., pore size distribution and connectivity), which makes the theoretical endeavour limiting to simplified model systems. The formation of fingering and its transition in these heterogeneous media are still an uncleared issue.

The early works describe the fingering phenomena mainly by the help of classical theories [9], including Hele-Shaw-Chouke theory [10], Buckley-Leverett theory [11], percolation theory [12], and Monte Carlo theory [13]. Later, fluid-fluid displacement has been further studied experimentally and numerically, especially in terms of flow in porous media [14]. Generally, capillary number *Ca* (the ratio of viscous forces to capillary forces) and viscosity ratio *M* (the ratio of the viscosity of defending phase to that of invading phase) are regarded as two major nondimensional parameters that control fingering patterns. Lenormand et al. [15] identified the fluid-fluid displacement patterns into three groups, i.e., viscous fingering, capillary fingering, and stable displacement, through combining capillary and viscous effects characterized by *Ca* and *M*, and originally provided a unified phase diagram. Since then, a series of modifications have been proposed based on this phase diagram [16-19]. However, the viscous and capillary effects are different on the characteristic time scales, especially after breakthrough during which the displacement is dominated by capillary effects and capillary rearrangement continues for a very long time, as suggested in Ref [45].

Further research suggests that the wettability, the spreading tendency of one fluid on a solid surface in the presence of another immiscible fluid, can also significantly affect the displacement patterns apart from fluid properties and flow conditions. It is generally quantified by the contact angle measured within the invading phase. The characteristics of wettability have been well established [20-26], with an understanding of the mechanism where the displacement pattern generally becomes more compact as an invading fluid is further wetted to the medium due to the cooperative pore filling effect. The fingering pattern changes extensively as a result of corner flow with the wettability increase of invading fluid.

Another important factor is topology of porous media. In contrast with the recent advances of wettability, the impact of porous structures on fingering patterns requires further research due to the complexity and diversity of porous media. Recent studies mainly focused on the heterogeneous porous media containing two groups, i.e., random and patterned structures. To construct random porous media, the particle size is generally assumed to follow a certain probability distribution, e.g. uniform distribution [27], power law distribution [28], normal distribution [29], and spatial-correlation Gaussian distribution [30], where fingering patterns change generally with disorder levels. Moreover, the interplay between the wettability and topological randomness has been reported recently [31, 32]. Additionally, various patterned structures have attracted considerable focus for controlling fingering patterns, e.g., Rabbani et al. [33] studied viscous fingering in ordered porous media with gradient pore sizes and such special structures were found to suppress viscous fingering to some extent. Zhang et al. [2] and Liu et al. [3] investigated liquid $CO_2$-water displacement in layered porous media with dual permeability. For miscible flow, Nijjer et al. [34] studied the effect of permeability heterogeneities and viscosity variations on displacement processes in layered porous media.

Besides the above two types of heterogeneity, porous media with hierarchical structures are also commonly encountered in natural and synthetic materials, such as fractured rock formations [35], dual-porosity media [36, 37], fibrous fabric materials [38] and silica monoliths [39]. In such materials, the microstructure can be considered as two or more interacting pore systems in different length scales, which collectively have a strong influence on fluid transfer properties. However, there are only limited research exploring the multiphase flow behaviours in such porous media to date, in particular, the effect of hierarchical structure on viscous fingering remains unexplored.

In this work, we investigate the hierarchically structured porous media and developing an analytical quantitative description of the fluid-fluid displacement modes in such materials. We first present a numerical model with two-level hierarchical porous structures to demonstrate how the micro-structure at different length scales, as well as the wettability condition, determine the fingering patterns during the displacement of one fluid within the pore space by a second immiscible fluid with a smaller viscosity. Specifically, we emphasis here on the viscous fingering regime, i.e., a hydrodynamic instability that occurs when a defending fluid is displaced by a less-viscous invading fluid, in a Hele-Shaw cell or porous medium. Meanwhile, we also present the evolution of fingering profiles due to the capillary rearrangement in a relative long duration at the post-breakthrough stages. The latter becomes relevant if there are two distinctive pore sizes presented in the media.

Our simulation results show that there are three fluid-fluid displacement modes, namely, "dendrite", "crossover", and "suppressing" modes, and the transition among these fingering modes can be controlled by changing the hierarchical topology and/or tuning the wettability condition. Furthermore, to quantitatively characterise fingering patterns in hierarchical porous media and predict the mode switch conditions, by combining driving pressures and interaction duration as two key factors, a scaling parameter (i.e., hierarchical number, $Hi$), is proposed. On this basis, a phase diagram is constructed for describing the observed displacement modes. Understanding of such fingering mode switch will pave the way for optimising design of many applications, such as, microfluidic devices and chemical reactors.

## 2. Numerical method

Navier-Stokes equations combined with the volume of fluid (VOF) method have been proved as an efficient way for pore-scale simulations of multiphase flow problems [40, 41], and the basic mathematical model for incompressible two-phase flow can be summed up as a system of equations, i.e., the continuity equation (1), phase fraction equation (2), and momentum equation (3), as

$$\nabla \cdot \mathbf{u} = 0, \tag{1}$$

$$\frac{\partial \phi}{\partial t} + \nabla \cdot (\phi \mathbf{u}) + \nabla \cdot \left[ \phi(1-\phi)\mathbf{u}_r \right] = 0, \tag{2}$$

$$\frac{\partial (\rho \mathbf{u})}{\partial t} + \nabla \cdot (\rho \mathbf{u}\mathbf{u}) - \nabla \cdot (\mu \nabla \mathbf{u}) - \nabla \mathbf{u} \cdot \nabla \mu = -\nabla p + F_c; \tag{3}$$

the boundary conditions as

$$\mathbf{u} \cdot \mathbf{n}\big|_{inlet} = v_{in}, \ \mathbf{u} \cdot \mathbf{\tau}\big|_{inlet} = 0, \ p\big|_{outlet} = p_{out}, \ \phi\big|_{inlet} = 1; \tag{4}$$

the initial conditions as

$$\mathbf{u}\big|_{t=0} = \mathbf{0}, \ p\big|_{t=0} = p_{out}, \ \phi\big|_{t=0} = 0, \tag{5}$$

where $\phi$ is the phase fraction of two fluids, i.e., $\phi=1$ for the invading phase and $\phi=0$ for the defending phase. $\mathbf{u}$ is the weighted average of velocity field shared by two fluids, i.e., $\mathbf{u} = \phi \mathbf{u}_{f1} + (1-\phi)\mathbf{u}_{f2}$, and $\mathbf{u}_r$ is the relative velocity, i.e., $\mathbf{u}_r = \mathbf{u}_{f1} - \mathbf{u}_{f2}$, $\rho$ and $\mu$ represent the weighted average of density and viscosity separately, i.e., $\rho = \phi \rho_{f1} + (1-\phi)\rho_{f2}$ and $\mu = \phi \mu_{f1} + (1-\phi)\mu_{f2}$, $p$ is the pressure. The capillary force, $F_c$, in Eq. (3), is defined as

$$F_c = \gamma \kappa \nabla \phi, \tag{6}$$

where $\gamma$ is the surface tension, and $\kappa$ is the mean curvature of the interface between two fluids and related to the interface normal $\mathbf{n}_\phi$,

$$\mathbf{n}_\phi = \frac{\nabla \phi}{|\nabla \phi|}, \tag{7}$$

$$\kappa = \nabla \cdot \mathbf{n}_\phi. \tag{8}$$

The wettability condition of solid boundaries, in this numerical scheme, can be considered by modifying the local normal to the interface contacting the solid boundaries, and specifically

$$\mathbf{n}_\phi = \mathbf{n}_s \cos(\theta) + \boldsymbol{\tau}_s \sin(\theta), \tag{9}$$

where $\theta$ is the contact angle; $\mathbf{n}_s$ and $\boldsymbol{\tau}_s$ are the normal and tangent vector to the solid boundary, separately. Here, the ideal solid surface is assumed in our simulation cases, and thus the hysteresis effect of contact angle is ignored. Computation of the equation set (1)-(3) and corresponding boundary conditions are performed by OpenFoam, an open source CFD toolbox [42], which adopts a cell centred finite volume method on a fixed unstructured numerical grid and utilizes the pressure implicit with splitting of operators (PISO) algorithm for the coupling between pressure and velocity in transient flows [43]. Before simulation, the sensitivity analysis including mesh size and time step has been conducted to guarantee the numerical accuracy and stability.

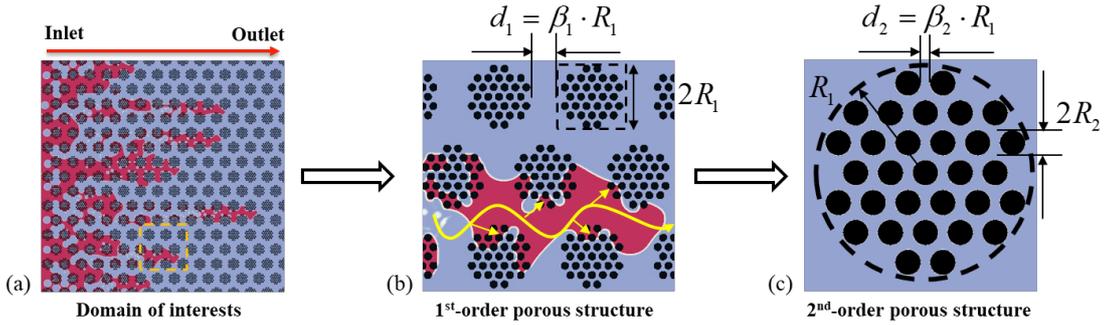

FIG. 1. Sketch of the numerical model of a two-level hierarchical porous medium. (a) Representative area of the numerical model with fluid displacement, in which the boundary conditions are set as velocity boundary at the left side, outflow boundary at the right side, and both bottom and top sides are set as periodic boundaries; (b) 1st-order porous structure with $R_1$ the radius of 1st-order obstacles and $d_1=\beta_1 \cdot R_1$ the 1st-order throat sizes; (c) 2nd-order porous structure with $d_2=\beta_2 \cdot R_1$ the 2nd-order throat sizes.

In this study, we consider the effect of the hierarchical structure on fingering patterns, and therefore fluid properties and flow conditions in all numerical cases are fixed except porous geometry and wettability. A schematic of the present numerical model is shown in FIG. 1. Our study is on a 2D domain without out-of-plane flow or gradients and it is initially saturated with the defending phase and then invading fluid is injected from the left. The left side is a uniformly inlet boundary with a fixed-value flow rate $v_{in}=1$ mm/s and the direction of the inlet velocity is perpendicular to the boundary, while the outlet boundary (right side) is set with a total pressure value $p_{out}=0$ Pa; the top and bottom sides are no-slip walls with contact angle of 90°. It is assumed that an invading fluid of viscosity $\eta_{in}=1\times10^{-3}$ Pa·s displaces a defending fluid of viscosity $\eta_{out}=1$ Pa·s, and density of both fluids equals $1\times10^3$ kg/m³; and the surface tension $\gamma$ between these two fluids equals 28.2 mN/m, resulting in the viscosity ratio of $M=\eta_{out}/\eta_{in}=1000$ and capillary number of $Ca=\eta_{out}v_{in}/\gamma=3.55\times10^{-2}$. This combination of $Ca$ and $M$ leads to viscous fingering occurring during fluid displacements and it is a typical scenario of water-oil system in practice [33]. In this work, in order to focus the topology effect on fluid displacements, we fixed the value of $Ca$ and $M$.

As for the geometry, both the 1st and 2nd obstacles are arranged on regular hexagonal grids, due to its structural homogeneity, in a self-affined manner. This geometry can be considered as the simplest representation of hierarchical porous media. Here, we choose the radius of 1st-order obstacles $R_1$ as a basic dimension parameter, and its value is set as 0.8 mm. On this basis, the throat size at both two levels are defined by introducing the corresponding factors, i.e., the 1st-order throat size $d_1$ equals $\beta_1 \cdot R_1$ and $\beta_1$ covers 0.54, 0.83 and 1.11, and thus the corresponding domain sizes are 30.1 mm×29.8 mm, 31.7 mm×31.3 mm and 33.3 mm ×32.9 mm. Although the domain size changes with the radius $R_1$, a sufficient number of the 1st-order obstacles have been included in a periodic structure, i.e., the domain size is significantly larger than the radius $R_1$, which usually guarantees to contain simultaneously multiple fingers. Thus, the present geometry can satisfy our requirement of capturing the specific modes of fingering. The 2$^{nd}$-order throat size $d_2$ equals $\beta_2 \cdot R_1$ and $\beta_2$ covers 0.075, 0.12, 0.14, 0.16, 0.18 and 0.20. In the present geometry, the radius $R_2$ is variable as to meet the requirement for the 2$^{nd}$-order porosity, with the relation of $\phi_{2nd} = 1 - n \cdot (R_2/R_1)^2$, where $n$ is the number of 2$^{nd}$-order obstacles. Notably, such design may introduce variation in the 2$^{nd}$-order pore throat curvature that determines the capillary effects during the fluid invasion [26, 44]. However, the effect of pore throat curvature has been captured in the later scaling analysis. Alternatively, one may fix the radius $R_2$ while varying the 2$^{nd}$-order arrangement. Additionally, the simulations are performed with a group of intrinsic contact angle $\theta$ (measured within the invading phase), i.e., 15°, 30°, 45°, 60°, 90°, and 120° to cover a wide range of wettability conditions. Note that in hierarchical porous media, the intrinsic contact angle alone does not determine the drainage and imbibition conditions, which will be further influenced by the pore geometry, as discussed later in Section 3.3.

Table 1 Geometry parameters for (a) 1$^{st}$-order and (b) 2$^{nd}$-order pore structures

(a)

| $\beta_1$ | $d_1$ (mm) | $R_1$ (mm) | $\phi_{1st}$ | $k_{1st}$ (mm$^2$) |
|---|---|---|---|---|
| 0.54 | 0.43 | 0.8 | 0.43 | 0.0114 |
| 0.83 | 0.66 | 0.8 | 0.55 | 0.0337 |
| 1.11 | 0.89 | 0.8 | 0.62 | 0.0698 |

(b)

| $\beta_2$ | $d_2$ (mm) | $R_2$ (mm) | $\phi_{2nd}$ | $k_{2nd}$ (mm$^2$) |
|---|---|---|---|---|
| 0.075 | 0.060 | 0.105 | 0.47 | 2.34×10$^{-4}$ |
| 0.12 | 0.096 | 0.087 | 0.63 | 8.28×10$^{-4}$ |
| 0.14 | 0.112 | 0.079 | 0.70 | 1.27×10$^{-3}$ |
| 0.16 | 0.128 | 0.070 | 0.76 | 1.85×10$^{-3}$ |
| 0.18 | 0.144 | 0.0626 | 0.81 | 2.62×10$^{-3}$ |
| 0.20 | 0.160 | 0.0556 | 0.85 | 3.60×10$^{-3}$ |

## 3. Results and discussion

The fluid-fluid displacement process in hierarchical porous media can be divided into pre- and post-breakthrough stages, split by breakthrough of the invading phase (i.e., the moment when the invading fluid phase reaches the outlet). The fingering patterns evolve rapidly during the pre-breakthrough stage since at the front of the invading phase, high pressure gradients occur and induce the nonuniform diffusion. Whereas, the pressure gradient decreases significantly when the invading fronts touch the outlet, and in the post-breakthrough stage, the overall pressure is inclined to be uniform, so the fingering evolutions proceed progressively slowly and tend to be stable. Here, we adopt a non-dimensional time $\tilde{t} = \dfrac{t}{d_1 \phi_{1st}/v_{in}}$ to estimate the duration. Compared with the time scale of viscous fingering (e.g., typically in seconds, i.e., $\tilde{t} \approx 10$) before breakthrough, the so-called capillary rearrangement during the post-breakthrough stage can last for a relative long time (e.g., about tens of minutes, i.e., $\tilde{t} > 200$), as suggested in Ref [45]. In this study, our study limits to the pre-breakthrough stage and early stage of capillary rearrangement.

## 3.1 Fingering modes

By employing the proposed numerical model, the multiphase flow simulation is performed in porous media with different hierarchical topology designs and wettability conditions. The simulation results show that the fluid-fluid flow pattern switches mainly among three fingering modes, i.e., supressing mode, crossover mode, and dendrite mode, corresponding to FIG. 2(a), (b) and (c), respectively. Noteworthily, these panels are parallelized in FIG. 2 only for demonstrating and comparing the main features of the three fingering patterns, so these three cases are chosen here only because they are representative for the corresponding group and other cases show the similar tendency with them.

To comprehensively quantify the different fingering pattern demonstrated in FIG. 2, four representative quantities are extracted from each simulation case: (I) displacement efficiency $F_e$, i.e., the ratio between the invasion fluid volume and the total fluid field volume, which is always adopted as an indicator during the oil collection [46], as shown in FIG. 3(a); (II) the effective fluid-fluid interface length $F_l$, i.e., the ratio between the fluid-fluid interface length (normalised by the domain width), obtained by Canny edge detection method [47], and the invasion fluid volume (normalised by the total volume), which often determine the chemical reaction rate [48], in FIG. 3(b); (III) fractal dimension $F_d$, which measures the roughness of the interface, computed through box-counting algorithm [49], in FIG. 3(c); and (IV) $2^{nd}$-order invasion ratio $F_i$, i.e., the ratio between the volume of invading phase inside $2^{nd}$-order porous zone and totoal volume of the pore space of $2^{nd}$-order structures, in FIG. 3(d). These four indices can macroscopically represent the fluid-fluid displacement effect and qualitatively distinguish the extent of fingering phenomena.

*Pre-breakthrough stage*

Before the breakthrough point, for these three cases, the front of invading phase advances towards the outlet quickly and viscous fingerings form. The dendrite mode is featured by the condition that the invasion fluid cannot infiltrate the second-order porous structure and thus the porous particles act like solid ones (see FIG. 2(c)), which is similar to the behaviour in single-scale porous media [13]. With the enhancement of capillary suction ability of the $2^{nd}$-order structures (reduce the intrinsic contact angle or throat size for $\theta < 90°$), as for the case in FIG. 2(a), the $2^{nd}$-order porous zone imbibes the invasion fluid with the evolution of fingerings. This type of pattern can be referred as the suppressing mode, which is also observed in the single-scale ordered porous media with gradient pore sizes [33]. These two fingering modes are bridged by a crossover zone, a typical crossover mode is given in FIG. 2(b), where the flow pattern presents similar with the dendrite mode at the beginning while the second-order porous structure are invaded at rather slower rate compared with the suppressing mode.

*Post-breakthrough stage*

After the invading front touches the outlet boundary, main pathways of invading phase are built, and the evolution of fingering patterns is driven by the capillary effects, especially the imbibition within $2^{nd}$-order porous structures. For the supressing mode, the fingering profile slightly changes and tends to be stable since the invasion within $2^{nd}$-order porous structures is mostly finished during pre-breakthrough stage, as clearly shown in FIG 3(d). Comparatively for the crossover mode, the $2^{nd}$-order invasion begins after breakthrough so that the fingering profile continues to evolve and then gradually grow to be stable. For the dendrite mode, due to the low capillary suction, $2^{nd}$-order invasion is obstructed and thus the fingering profile keep almost unchanged once the invading fluid reach the outlet.

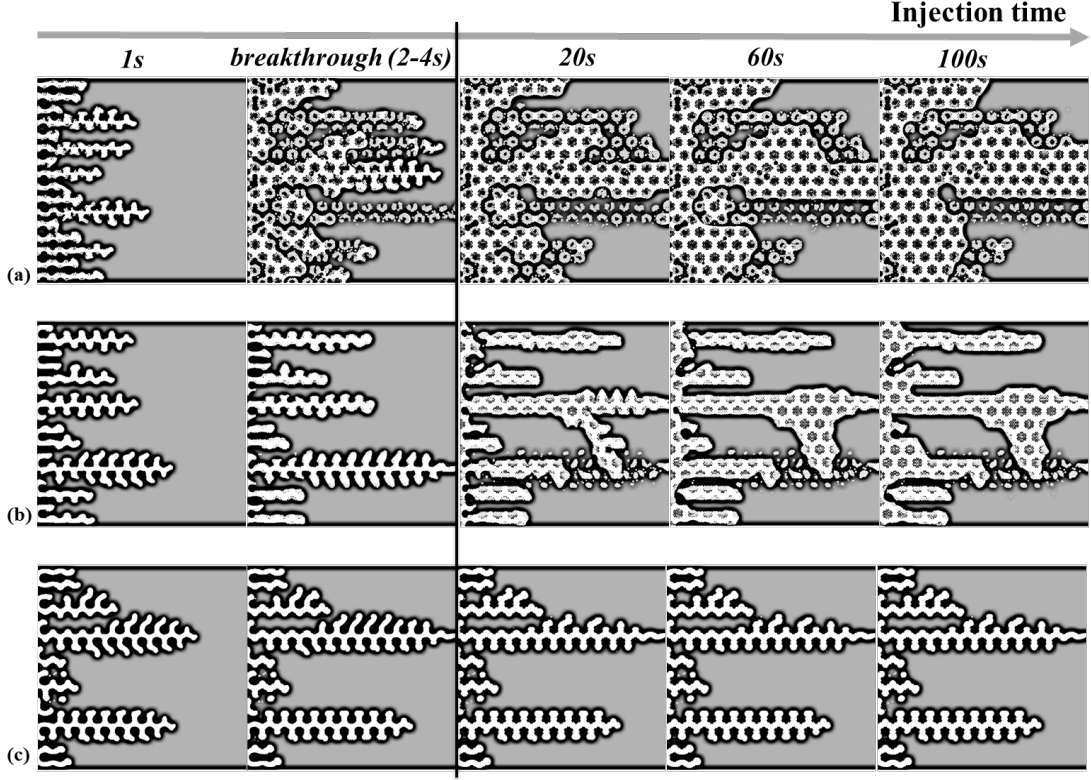

FIG. 2. Fingering patterns with three different modes: (a) suppressing mode ($\beta_1=0.83$, $\beta_2=0.075$, $\theta=15°$) with breakthrough time of approximately 4 sec; (b) crossover mode ($\beta_1=1.11$, $\beta_2=0.14$, $\theta=45°$) with breakthrough time of 2 sec; (c) dendrite mode ($\beta_1=0.83$, $\beta_2=0.14$, $\theta=120°$) with breakthrough time of 2 sec.

Generally, the quantitative characteristics of the three fingering modes given in FIG. 3 are overlapped to a certain extent at the pre-breakthrough stage, while they can be clearly distinguished by the quantitative characteristics at 20s, a typical moment of the early capillary rearrangement. In more quantitative terms, the values of $F_e$, $F_l$, $F_d$ and $F_i$ for the suppressing mode is larger than the crossover mode and then the dendrite mode with noticable differences. Furthermore, at 20s the gap among these three modes are large enough for distinguishing while at breakthrough point, some indices like $F_e$ and $F_d$ are overlapped for different modes. Thus, we take 20s result as representative of a displacement process for the following analysis.

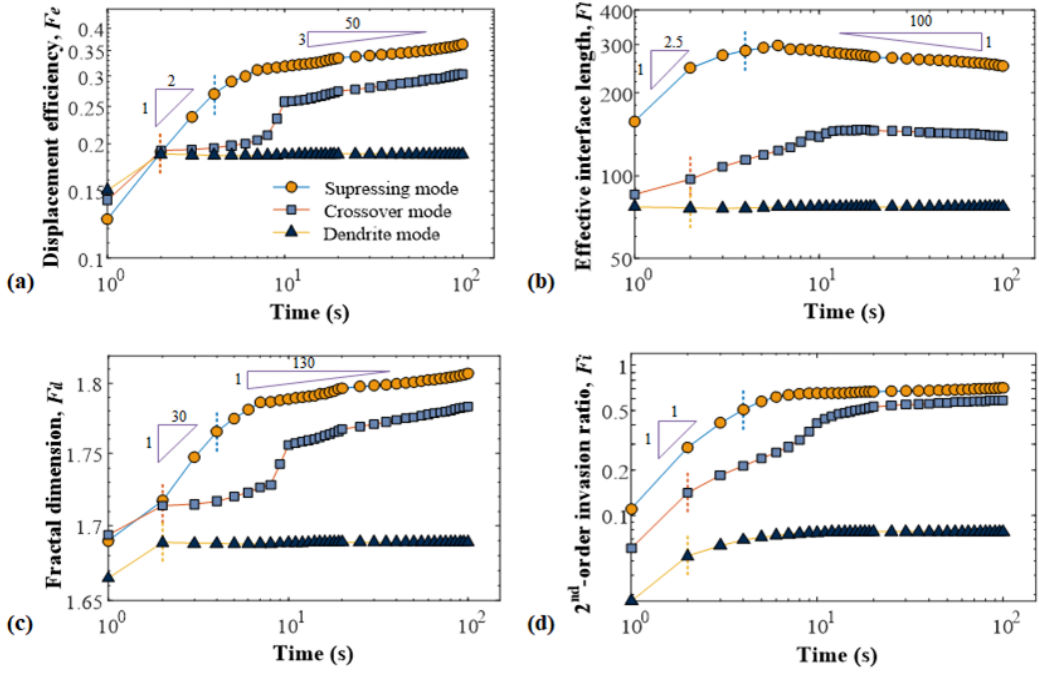

FIG. 3. Fingering indices as a function of injection time, including (a) displacement efficiency $F_e$, (b) effective interface length $F_l$, (c) fractal dimension $F_d$, and (d) 2$^{nd}$-order invasion ratio $F_i$, and the cases corresponding to the dendrite, crossover and suppressing modes in FIG. 2 and the corresponding breakthrough time is marked by a short dash line.

## 3.2 Phase diagram

The combined effects of wettability conditions and hierarchical topological feature (more specifically, the interaction between the 1$^{st}$-order and 2$^{nd}$-order porous structures) on the transitions of fingering modes are summarized as the phase diagrams shown in FIG. 4. As can be seen in FIG. 4(a), with a fixed size of the 1$^{st}$-order throat size $d_1$, the fingering pattern transforms from the supressing mode to the dendrite mode as obstacle surfaces become gradually hydrophobic (the contact angle varies from 15° to 120°) for invasion fluid. Furthermore, the phase boundary between crossover mode and dendrite mode mainly depends on the wettability conditions, and it is less sensitive to the changes of the 2$^{nd}$-order pore geometry. For instance, the fingering pattern will transit to the dendrite mode with the contact angle beyond 90° in this study. Indeed, there would be difference among suppressing cases since geometries are different after all. However, such difference should be slight, because the pre- and post- breakthrough behaviours are controlled by the capillarity and viscosity while the capillary number and viscosity ratio are fixed in this study since we focus on the geometric effect. The $2^{nd}$-order throat size $d_2$ controls the porosity at the lower scale (as well as the overall porosity) and keeps the apparent $1^{st}$-order porosity constant. With increasing the $2^{nd}$-order throat size $d_2$, the capillary pressure that drives invasion fluid to spontaneously fill the $2^{nd}$-oder porous structures decreases, while the permeability of $2^{nd}$-order porous structures increases resulting in low flow resistance. Therefore, the crossover region broadens whereas the phase boundary between suppressing mode and crossover mode becomes ambiguous to some extent.

To further uncover the hierarchical effect on fingering patterns, another phase diagram between wettability conditions and $1^{st}$-order throat size $d_1$ (with the fixed $2^{nd}$-order throat size $d_2 = 0.14 R_1$) is provided, as shown in FIG. 4(b). The $1^{st}$-order throat size $d_1$ controls the $1^{st}$-order porosity $\phi_1$ and permeability $k_{1st}$. According to Darcy's law, the pressure gradient is inversely proportional to the $1^{st}$-order permeability $k_{1st}$, since the inlet flow rate is fixed during the whole displacement process. Specifically, the pressure gradient and the corresponding external entering pressure that drives the invasion fluid to infiltrate porous obstacles decreases with the increment of $1^{st}$-order throat size. Thus, the suppressing mode occurs with rather smaller $1^{st}$-order throat size. For wettability, similarly with FIG. 4(a), the dendrite mode is found for all the cases with contact angle no less than 90°, and the suppressing mode can only be found for the cases with $\theta < 30°$.

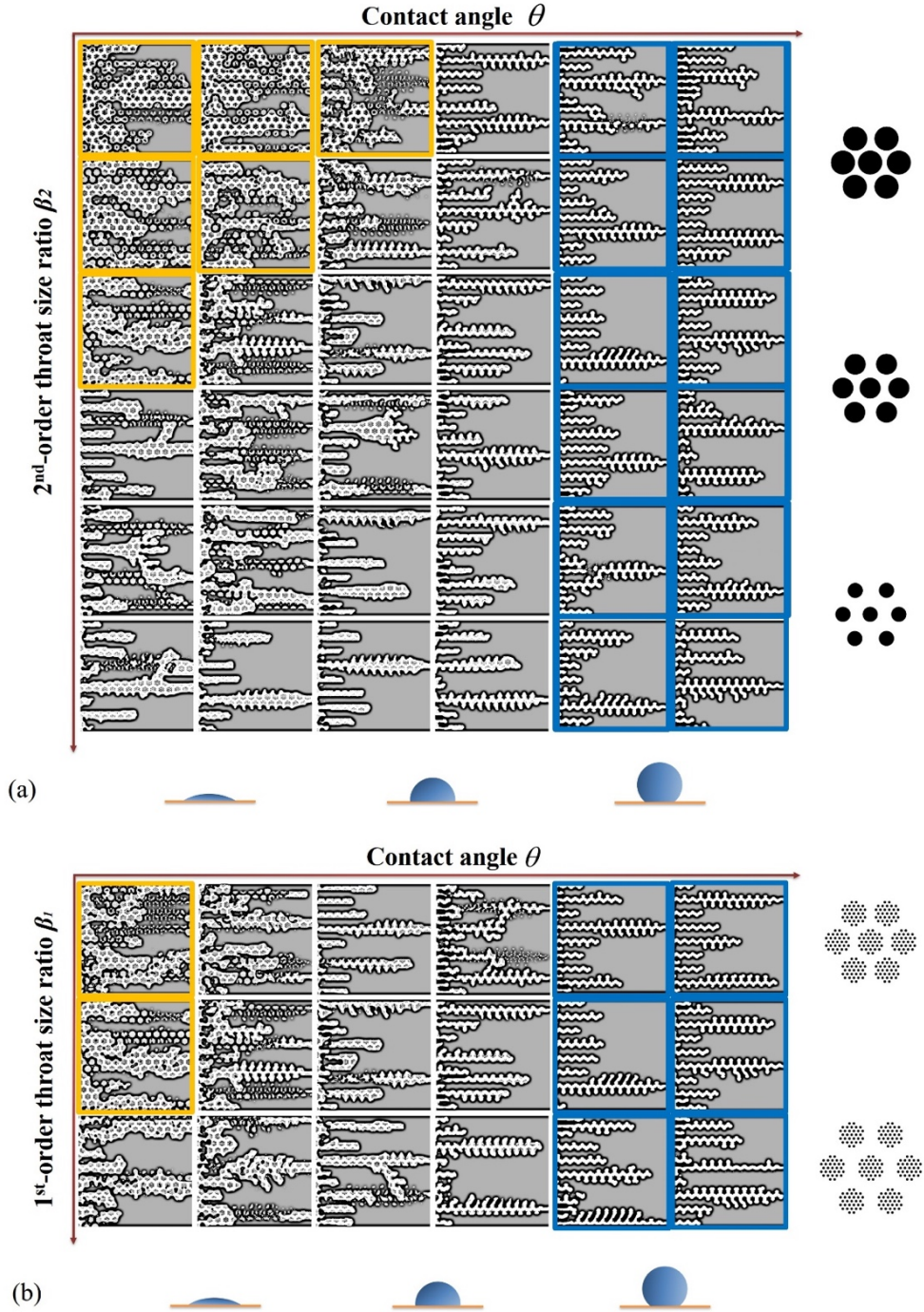

FIG. 4. Phase diagram of fingering patterns in hierarchical porous media at different contact angles measured within the invading phase (i.e., $\theta$ = 15°, 30°, 45°, 60°, 90°, and 120°) and (a) $2^{nd}$-order throat sizes (i.e., $\beta_2$ = 0.075, 0.12, 0.14, 0.16, 0.18 and 0.20) with fixed $1^{st}$-order throat size of $0.83R_1$; (b) $1^{st}$-order throat sizes (i.e., $\beta_1$ = 0.54, 0.83, and 1.11) with fixed $2^{nd}$-order throat size of $0.14R_1$. Among those, ▫ denotes suppressing mode and ▫ dendrite mode, while the rest are crossover mode. All results are taken at t=20s.

## 3.3 Dimensionless analysis

Based on the above analysis of phase diagrams, it suggests that the transition of fingering patterns is controlled by driving pressure and interaction duration. The first determines a thresholding process for invading fluid to enter the 2$^{nd}$-order pore space, while the latter describes the competition of fluid flow at two different pore scales.

**a) Driving pressure**

To quantitatively depict the pressure that drives the invasion fluid into the 2$^{nd}$-oder porous structures, the representative element shown in FIG. 5(a) is adopted here. The capillary pressure $P_c$, as a function of filling angle $\alpha$, can be derived as [50]

$$P_c = \frac{\gamma}{d_2} \frac{\cos(\theta - \alpha)}{1 + 2R_2/d_2 (1 - \cos(\alpha))}, \tag{10}$$

where $\alpha$ is the filling angle. For a given wettability and geometry condition, the capillary pressure $P_c$ reaches two extreme values, i.e., $P_{c,\min}$ and $P_{c,\max}$ within $\alpha \in \left[-\frac{\pi}{2}, \frac{\pi}{2}\right]$ at $\alpha^{\min}$ and $\alpha^{\max}$ respectively, and they can be expressed as

$$\alpha^{\min} = \theta + \sin^{-1}\left(\sin(\theta)/(1 + r_2/R_2)\right) - \pi, \tag{11}$$

and

$$\alpha^{\max} = \theta - \sin^{-1}\left(\sin(\theta)/(1 + r_2/R_2)\right). \tag{12}$$

Besides, another key point is the critical filling angle $\alpha^0 = \theta - \pi/2$ at which $P_c$ equals zero. A general infiltration process is separated into two parts by $\alpha^0$. Firstly, the fluid-fluid interface is forced to move across the 2$^{nd}$-order throat since $P_c < 0$ when $\alpha < \alpha^0$, named the "inhibited process"; then, the interface movement becomes spontaneous when $\alpha > \alpha^0$ and thus $P_c > 0$, named the "spontaneous process". During the inhibited process, the infiltration is driven by the external pressure $P_{ext}$, and specifically for supressing mode and crossover mode, it is basically required that $P_{ext}$ is large enough to break through the barrier value $P_{c,\min}$ and start the spontaneous process. To depict this effect, a dimensionless number $R_p$ is defined here as

$$R_p = \frac{P_{ext}}{P_{c,\min}}. \tag{13}$$

Obviously, the larger $R_p$ is, the more likely the fingering pattern switches to the suppressing mode. In this study, $P_{ext}$ can be estimated according to Darcy's law as

$$P_{ext} = \frac{v_{in}\eta_{in}}{k_{1st}}d_1, \tag{14}$$

where $k_{1st}$ is the permeability of the 1st-order porous structure.

The relationship between the controlling parameters, i.e. $P_{c,\min}$, $P_{c,\max}$ and $\alpha^0$, and contact angle $\theta$ can be referred to FIG. 5(b). Generally, with the increment of contact angle, $P_{c,\min}$ and $P_{c,\max}$ both decrease while the critical filling angle $\alpha^0$ increases. For drainage cases ($\theta > 90°$), $\alpha^0 > 0$, and thus the "inhibited process" is dominating; for imbibition cases ($\theta < 90°$), the pressure barrier $P_{c,\min}$ is rather small and the "spontaneous process" is governing. In the latter case, the invading phase may be more inclined to infiltrate the 2nd-order porous structures as imbibition.

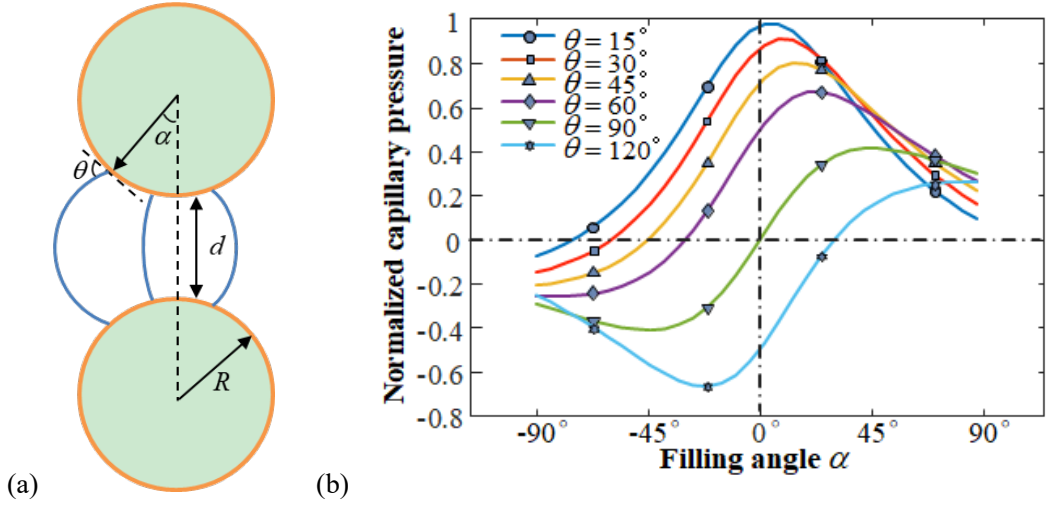

(a)                  (b)

FIG. 5. (a) Two-sphere model for calculating the capillary pressure as invading fluid moves through the throat where $d$ is the throat size, $\alpha$ is the filling angle, and $\theta$ is the contact angle; (b) Filling angle $\alpha$ vs. normalized capillary pressure, i.e., the ratio of capillary pressure $P_c$ in Eq. 8 to $\gamma/d_2$.

b) **Interaction duration**

The main difference between the suppressing mode and the crossover mode is the infiltration rate, and to further reflect such difference in detail, we propose another dimensionless number $R_T$, which is the ratio between two characteristic time scales at the respective pore scales. Specifically, $T_{1st}$ is the time during which the invading fluid flows through the 1st-order throat,

$$T_{1st} = \frac{\phi_{1st} d_1}{v_{in}}, \tag{15}$$

where $\phi_{1st}$ denotes the first-order porosity; $T_{2nd}$ is the time for the invasion fluid to infiltrate 2nd-order porous structure by capillary suction, and it can be derived from the following modified Darcy's law [51],

$$\phi_{2nd} \frac{\partial y}{\partial t} = \frac{k_{2nd}}{\eta_{in}} \frac{P_c^*}{y}, \tag{16}$$

where $y$ is the suction distance; $\phi_{2nd}$, the second-order porosity; $k_{2nd}$, the permeability of the $2^{nd}$-order porous structure, is a function of $\phi_{2nd}$ and $d_2$, i.e., $k_{2nd} = d_2^2 \cdot \dfrac{\phi_{2nd}^m}{(1-\phi_{2nd})^n}$ (with $m = 0.76$ and $n = 0.24$ for the studied geometry) which can be referred in the *Supplementary Material*; and $P_c^*$ is a representative value of capillary pressure and can be estimated from Eq. (10) as

$$P_c^* = \dfrac{\int_{\alpha_0}^{\pi/2} P_c(\alpha) d\alpha}{\pi}. \tag{17}$$

Assuming the characteristic infiltration distance as $d_2$, $T_{2nd}$ is solved as

$$T_{2nd} = \dfrac{\eta_{in} \phi_{2nd} d_2^2}{2 k_{2nd} P_c^*}. \tag{18}$$

Thus, $R_T$ is expressed as

$$R_T = \dfrac{T_{1st}}{T_{2nd}} = 2 \dfrac{\phi_{1st} k_{2nd} P_c^* d_1}{\phi_{2nd} v_{in} \eta_{in} d_2^2}. \tag{19}$$

From Eq. (19), increasing $R_T$ means that it takes less time for the invading phase to infiltrate into the $2^{nd}$-order porous structures with fixed $1^{st}$-order flow conditions, and thus it is more posssible for fingering pattern transiting to suppressing mode with larger $R_T$. The parameter can be used to quantify the effect of the interaction between two-scale structure feature on fluid flow behaviours.

Multiplying both two dimensionless number $R_p$ and $R_T$, a complete dimensionless number, namely *hierarchical number* Hi, to comprehensively characterise the fingering patterns in hierarchical porous media, is defined here as

$$\text{Hi} = R_p R_T = 2 \dfrac{P_c^*}{P_{c,\min}} \dfrac{\phi_{1st}}{\phi_{2nd}} \dfrac{k_{2nd}}{k_{1st}} \left(\dfrac{d_1}{d_2}\right)^2. \tag{20}$$

Above all, this hierarchical number combines the two effects, i.e., driving pressure and interaction duration. So, when Hi increases, the fingering pattern is expected to transit from dendrite mode to suppressing mode accordingly.

## 3.4 Validation and correlation analysis

To verify the validity of the proposed hierarchical number dimensionless parameter, i.e., hierarchical number Hi, the cloud charts and contours of Hi, as a function of the wettability conditions (i.e., contact angle) and the geometric parameters (i.e., relative throat size $\beta_1$ and $\beta_2$), are presented in FIG. 6. The cases in FIG. 3 are also collected and displayed on the cloud map, with different type of symbols to represent different modes. It is clearly seen that the different fingering modes can be uniformly distinguish by the value of Hi for both two cases in FIG. 6(a) and 6(b). More specifically, the dendrite mode occurs when $Hi \geq 3.9$, while the supressing mode occurs when $Hi \leq 0.9$. At the mean while, in the condition of $0.9 < Hi < 3.9$, the crossover mode occupies the transition region.

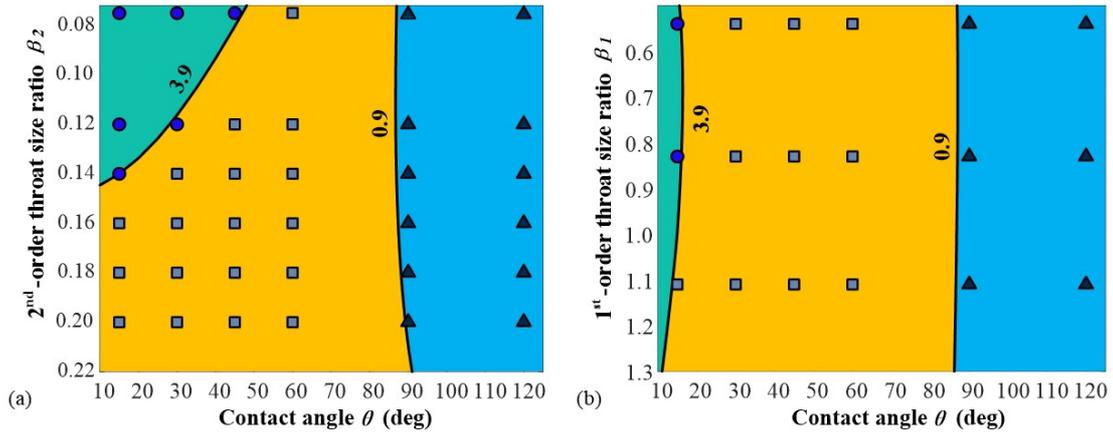

FIG. 6. Cloud chart and contour lines of hierarchical number Hi as a function of contact angle and (a) 2$^{nd}$-order relative throat size $\beta_2$ and (b) 1$^{st}$-order relative throat size $\beta_1$; and scatter ○ denotes supressing mode, ■ crossover mode, and ▲ dendrite mode, which correspond to that in FIG. 2.

Furthermore, the fingering indices ($F_e$, $F_l$, $F_d$ and $F_i$) of all the cases in Fig. 3 is calculated and presented as a function of the proposed hierarchical number $\mathrm{Hi}$ in Fig. 7. It is clearly seen that all these four indices demonstrate strong positive correlations with the hierarchical number, with the correlation coefficient R ranging from 0.83 to 0.92. Moreover, three clusters of the representative quantities, corresponding three fingering patterns, distribute in three distinct $\mathrm{Hi}$ regions respectively, i.e., the dendrite mode occurs when $\mathrm{Hi} \leq 0.9$, suppressing mode when $\mathrm{Hi} \geq 3.9$, and the middle section belongs mainly to crossover mode almost since there are some overlapping areas between the suppressing region and the crossover region.

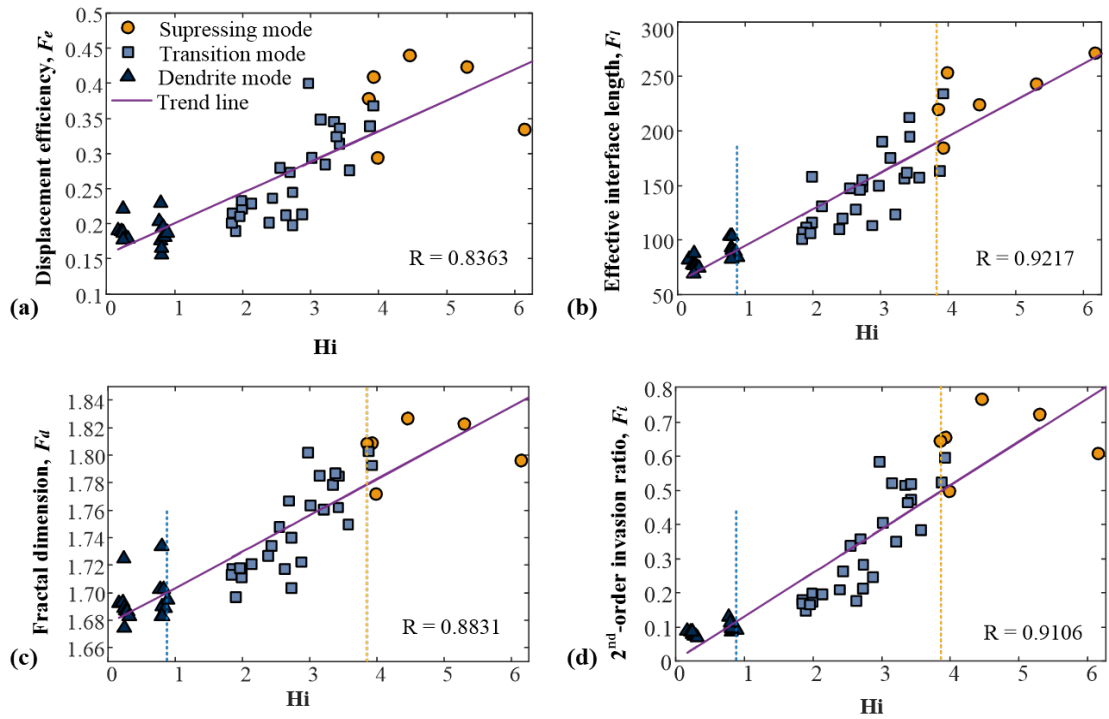

FIG. 7. The correlation between hierarchical number Hi and (a) displacement efficiency $F_e$, (b) effective interface length $F_l$, (c) fractal dimension $F_d$, and (d) 2$^{\mathrm{nd}}$-order invasion ratio $F_i$. All results are taken at t=20s.

In summary, the proposed dimensionless parameter, i.e., *hierarchical number,* $\mathrm{Hi}$, can reflect the effects of geometric feature of porous structures and wettability conditions on the fluid-fluid behaviours in hierarchical porous media, and the corresponding conditions of the transition of fingering patterns can be quantitatively predicted by $Hi$. Following this parameter, the fingering behaviours can be evaluated uniformly and further optimization.

## 4. Conclusion

In this work, we investigated the fingering phenomena in hierarchical porous media and the impacts of topology and wettability on fingering patterns in hierarchical porous media are studied through pore-scale numerical simulations. The fingering processes in hierarchical pore structures demonstrate two distinct regimes of fluid displacement driven by viscous and capillary effects at two difference time scales. The main contributions of this work are listed as follows, (1) three different fluid-fluid displacement modes, namely, the dendrite, crossover and suppressing modes, are observed in the hierarchical porous media considering the combined effects of wettability and geometric structural features; (2) a phase diagram is provided to describe the transition among the fingering patterns, depending on the wettability and hierarchical features; (3) and a new dimensionless number $Hi$ is proposed by combining main driving mechanisms to predict this transition of fingering modes. Especially, according to the analysis, we propose a possible way to suppress the immiscible fingering by adjusting the pore topology of porous media.

Here, the values of $M$ and $Ca$ are fixed for the main purpose of highlighting the dependency on multiscale pore topology under the condition of viscous fingering, as one of the key aspects of our current study. Indeed, different combinations of $M$ and $Ca$ may introduce variability of the exact location of phase boundaries. However, the main working principle of the analyses presented in this study can still employ, since the proposed dimensionless number $Hi$ captures the fundamental mechanisms governing the flow at different length scales. Besides, since the proposed scaling analysis only requires involves macroscopic geometry parameters, i.e. permeability and porosity, our scaling analysis may be extended to other types of hierarchical pore structures.

Moreover, the immiscible displacement in hierarchical porous media may share some similarities with the miscible displacement during which specifically the two main driving factors are the diffusion and advection for miscible flows and dimensionless number Pe, a ratio of advection term and diffusion term, control the miscible fingering patterns [52]. Similarly, the immiscible fingering in hierarchical porous media is also controlled by two components, i.e., the Darcy flow in $1^{st}$-order fluid paths and the infiltration in $2^{nd}$-order pore spaces. Future studies should explore the link between modes of immiscible and miscible displacement.

This study about the mechanism of the fingering mode transition warrants an effective method to tailor the viscous fingering by adjusting properties at different scales, such as permeability and porosity. The conclusion of this research may contribute to several potential engineering applications related to the optimisation of microfluidic device, agricultural irrigation and fuel cells.

**Acknowledgements**

This work was financially supported by Australian Research Council (Projects DP170102886) and The University of Sydney SOAR Fellowship. M.L. acknowledges the support of a Newton International Fellowship from the Royal Society. This research was undertaken with the assistance of resources and services from the HPC service at The University of Sydney. The authors would like to thank the reviewers for their generous and helpful comments.